\documentclass[aps,prl,twocolumn,groupedaddress]{revtex4}
\usepackage{epsfig}
\usepackage{graphicx}
\usepackage{amsmath}

\newcommand{\beq}{\begin{eqnarray}}
\newcommand{\eeq}{\end{eqnarray}}

\begin{document}

\title{Magnetic Seed Field Generation from Electroweak Bubble Collisions, with Bubble Walls of Finite Thickness}

\author{
Trevor Stevens\\
Department of Physics, West Virginia Wesleyan College, Buckhannon, WV 26201\\
Mikkel B. Johnson \\
Los Alamos National Laboratory, Los Alamos, NM 87545
}

\affiliation{}

\date{\today}

\begin{abstract}
Building on earlier work, we develop an equation-of-motion method
for calculating magnetic seed fields generated from currents arising from charged $W^\pm$ fields in bubble collisions during a first-order primordial electroweak phase transition allowed in some proposed extensions of the Standard Model.  The novel feature of our work is that it takes into account, for the first time, the dynamics of the bubble walls in such collisions. We conclude that for bubbles with sufficiently thin surfaces the magnetic seed fields may be comparable to, or larger than, those found in earlier work.  Thus, our results strengthen the conclusions of previous studies that cosmic magnetic fields observed today may originate from seeds created during the electroweak phase transition, and consequently that these fields may offer a clue relevant to extensions of the Standard Model. 
\end{abstract}

\pacs{ } 

\maketitle
PACS Indices:12.38.Lg,12.38.Mh,98.80.Cq,98.80Hw



\section{Introduction}

Explaining the origin of galactic and extra-galactic magnetic fields remains an outstanding problem in cosmology. 
A common approach is to view these fields as having arisen from magnetic seed fields created in the early universe, possibly during the electroweak phase transition.
Several papers \cite{kibble,ahonen,copeland} have shown how magnetic fields can arise from the equilibration of the Higgs phase within the collision region of the expanding electroweak bubbles in first-order phase transitions. First-order phase transitions are not possible for the Standard Model~\cite{klrs} but may be allowed~\cite{laine,cline,losada} in certain minimal extensions of the Standard Model (MSSM).

These early studies of magnetic seed formulation were formulated in the context of a simple Abelian model. 
In \cite{mv} we have proposed an alternative and, we believe, a more natural mechanism, which is that the charged $W$-fields are the physical origin of the electromagnetic currents creating the magnetic seed fields.  We found that the magnetic seed fields generated by this mechanism were of a magnitude comparable to those found in the Abelian model.

In \cite{mv} we obtain our results by solving equations of  motion (EOM) for the charged $W$-fields of the MSSM for bubbles that have collided, 
\beq
\label{eomw12}
\partial^2w^a_\nu-\partial_\nu\partial\cdot w^a
+m^2w^a_\nu=0~,
\eeq
where $a=(1,2)$ are the fields of the charged $W$ gauge fields and $m^2=\rho_0^2g^2/2$ is the square of its mass within the bubbles. Our method of solution followed previous 
studies with jump boundary conditions imposed on the $w_z$ field at the moment of collision. In this approach, the divergence $\partial\cdot w^a$ vanishes, a result that follows from taking the divergence of both sides of (\ref{eomw12}), 
\begin{equation}
\partial^{\nu}m^{2}w_{\nu}=m^2 \partial^{\nu}w_{\nu}=0
\end{equation} 
and taking $m(x)$ to be constant within the bubbles.

In the present work, we employ a generalization~\cite{thesis,stevens1} of Ref.~\cite{mv} that evolves the collision from bubbles that are initially separated to estimate, using representative boundary conditions, the importance of bubble surface dynamics on magnetic field creation.  In this case, $m$ can no longer be taken constant in $x$, and the divergence of $w$ is now expressed in terms of $m(x)$ as an {\it auxiliary condition}, 
\begin{equation}
\label{divvy}
\partial\cdot w+\frac{1}{m^2}w_{\nu}\partial^{\nu}m^2=0~,
\end{equation}
which the solutions of the EOM must 
the divergence of $w_\nu$ in the EOM represents
a coupling to the bubble wall that cannot be neglected as a contribution to the magnetic field. 
When surface is taken into account, solving Eq.~(\ref{eomw12}) subject to Eq.~(\ref{divvy}) requires a new approach, and the main difficulty in 
implementing it is having to solve the field equations numerically, in contrast to Ref.~\cite{mv}. We note in passing that for the case of bubbles with infinitely thin walls, $\partial m/\partial r$ becomes a delta function at the bubble surface, and consequently the $W$-fields (and hence the magnetic fields) may become singular in this limit~\cite{stevens2}.

Once the EOM have been solved for the $W$ fields, the electromagnetic current may be calculated~\cite{mv,thesis,stevens1} from the expression
\begin{equation}
\label{current}
4\pi j_{\nu}=G\epsilon^{ab3}\left(w_{\nu}^b (\partial\cdot w^a)-w_\mu^a \partial_{\nu} w^{\mu b}
+2w_{\mu}^a \partial^{\mu} w_{\nu}^b\right)
\end{equation}
where 
\begin{equation}
G=\frac{gg'}{\sqrt{g^2+g'^2}}~.
\end{equation}

As in~\cite{mv,thesis,stevens1}, the current will vanish if $w_{\nu}^b(x) \propto w_{\nu}^a(x)$. To ensure a non-vanishing current in the present work,
we choose boundary conditions at a time $t=t_0=0$ such that
the fields $w_{z}^b(t_0,\vec x) = w_{z}^a(t_0,\vec x)$ in one bubble, referred to as boundary condition II (BCII), and $w_{z}^b(t_0,\vec x) =- w_{z}^a(t_0,\vec x)$ in the other (BCI) with the values of $w_\nu(t_0,\vec x)$ constant in each bubble as in Ref.~\cite{mv}. Notice that fixing the boundary condition on $w_z$ in this way follows Ref.~\cite{mv}, where the $w_z$-field was taken
to be a step function at $t_0.$ Initial conditions for the other fields 
needed in our present work that result from these boundary conditions and the auxiliary condition are discussed below.

Other choices of boundary conditions for $w_z$ may also be envisioned,
and in principle all these should be averaged over to find the magnetic field. However, BCI and BCII are representative boundary consitions~\cite{mv} and are 
thus sufficient for our present 
purposes.

\section{Bubble Collision in Cylindrical Coordinates}

To study the case of two colliding bubbles, we will use the axial symmetry about the $x_1$ and $x_2$ axes to write the $w$ vector fields in cylindrical coordinates as:
\begin{equation}
w_{\nu} = \left\{ 
\begin{array}{ll}
         w_0, & \mbox{$\nu$=$0$};\\
         wx_{\nu} & \mbox{$\nu$=$1,2$};\\
        w_z & \mbox{$\nu$=$3$}.
\end{array} 
\right. 
\end{equation}

In cylindrical coordinates, the auxiliary condition becomes:
\begin{eqnarray}
\label{div}
\partial\cdot w(x) &=& -2w_{0}\left(\frac{1}{m}\right)\frac{\partial m}{\partial t}
-2rw\left(\frac{1}{m}\right)\frac{\partial m}{\partial r}\\ \nonumber
& &+2w_{z}\left(\frac{1}{m}\right)\frac{\partial m}{\partial z}~. \nonumber
\end{eqnarray}
Then the equations for the $W$-fields may then be written
\beq
\label{eom1}
\frac{\partial^2 w_0}{\partial t^2}&-&\frac{\partial^2 w_0}{\partial 
r^2}
-\frac{1}{r}\frac{\partial w_0}{\partial r}-\frac{\partial^2 
w_0}{\partial z^2}+m^2 w_0 \nonumber \\
&-&\frac{\partial}{\partial t}         \partial\cdot w(x) = 0
\eeq
\beq
\label{eom2}
\frac{\partial^2 w}{\partial t^2}&-&\frac{\partial^2 w}{\partial r^2}
-\frac{3}{r}\frac{\partial w}{\partial r}-\frac{\partial^2 w}{\partial z^2}
+m^2 w \nonumber \\
&+&\frac{1}{r}\frac{\partial}{\partial r}    \partial\cdot w(x) =0
\eeq
\beq
\label{eom3}
\frac{\partial^2 w_z}{\partial t^2}&-&\frac{\partial^2 w_z}{\partial 
r^2}
-\frac{1}{r}\frac{\partial w_z}{\partial r}-\frac{\partial^2 
w_z}{\partial z^2}+m^2 w_z \nonumber \\
&-&\frac{\partial}{\partial z}        \partial\cdot w(x) = 0~,
\eeq
using $\partial\cdot w(x)$ given in Eq.~(\ref{div}).
It is these equations which we solve to study the evolution of the $W$-fields in the case of two colliding electroweak bubbles. For the calculations presented in this paper, the equations are solved in {\em Mathematica}~\cite{MATH} using the built-in NDSolve function.

\subsection{Initial Conditions for Bubble Nucleation}

The auxiliary condition is maintained for all time if it (and its time derivative) 
is satisfied at an initial time $t_0$. This leads to initial conditions, or constraints, among the fields and their time derivatives at $t_0$~\cite{stevens1}. In addition to $w_z(t_0,\vec x)$, we are also free to choose initial values for $w_0(t_0,\vec x)$, $w(t_0,\vec x)$ and $\partial w_z(t_0,\vec x)/\partial t$; for simplicity, we take the latter to vanish.  The auxiliary condition, Eq.~(\ref{divvy}), then gives, in cylindrical coordinates,
\begin{eqnarray}
\label{w_div2}
\frac{\partial w_0(t_0,\vec x)}{\partial t}-\frac{\partial w_z(t_0,\vec x)}{\partial z} =
2 \frac{ w_{z}(t_0,\vec x)}{m(t_0,\vec x)}\frac{\partial m(t_0\vec x)}{\partial z}~.
\end{eqnarray}

The initial condition for the time derivative of $w^a_0(t_0,\vec x)$ follows directly from Eq.~(\ref{w_div2}) and the boundary conditions BCI and BCII on $w_z(t_0,\vec x)$.  These boundary conditions are satisfied with $w_z(t_0,\vec x)$ proportional to $\pm m(t_0,\vec x)$ when the colliding bubbles are well separated, which in turn may be satisfied if
\begin{equation}
\label{wzform}
\frac{\partial w_z(t_0,\vec x)}{\partial z}= 
\frac{ w_{z}(t_0,\vec x)}{m(t_0,\vec x)}\frac{\partial m(t_0\vec x)}{\partial z}~.
\end{equation}
Combining Eqs.~(\ref{w_div2},\ref{wzform}), the time derivative of $w_0$ at $t_0$ takes the form
\begin{equation}
\label{initcond}
\frac{\partial w_0(t_0,\vec x)}{\partial t}=3\frac{\partial w_z(t_0,\vec x)}{\partial z}~.
\end{equation}

\begin{figure}[h!]
\includegraphics{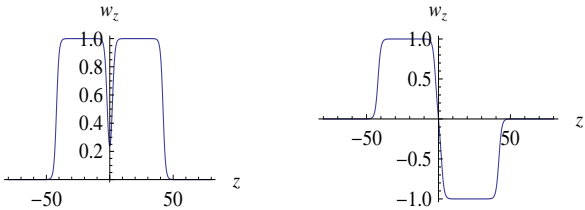}
\vspace{4.5cm}
\caption{Profile of initial conditions for $w_z$}
\label{subdpic}
\end{figure}

The profiles of $w_z(t_0,\vec x)$ for BCI and BCII are shown in Fig.~\ref{subdpic}.  It is seen that the bubbles collide shortly after nucleation with initial radii of $r_n=20$ in units of the inverse $W$ mass.  Corrections are required for the slight overlap of the bubbles and for fact that other bubbles give rise to an average scalar field, but these are small effects~\cite{stevens1} and our calculated magnetic field should be reasonably accurate without them.

The value of $w_z(t_0,\vec x)$ at $t=t_0$ is fixed here
by normalizing the $W$ fields inside the bubble as in Ref.~\cite{mv} to give a reasonable number of $W^\pm$-bosons
inside the bubbles under certain assumptions about the thermal conditions. By choosing the normalization in this way in both calculations, a direct comparison with the results 
of Ref.~\cite{mv} becomes more meaningful.

\subsection{Functional Form of the Scalar Field}

To solve the EOM in Eqs.(\ref{eom1},\ref{eom2},\ref{eom3}), we need the functional 
form of $m(x)$ as a function of time. 
For a single bubble, $m(x)$ is proportional to the magnitude of the scalar field $\rho(x)$, which at zero temperature is simply the analytic continuation of the bounce solution to the Coleman equation~\cite{coleman} 
\begin{equation}
\label{eomf}
\partial^2\rho(x)+\rho(x)\frac{\partial V}
{\partial \rho(x)^2}=0
\end{equation}
in Euclidean space, where the bubble walls expand smoothly and retain the functional form of the bounce. At nucleation, we assume the scalar field is a simplified version of the bounce solution given in Eq.~(6) of Ref.~\cite{copeland} with $\eta\sqrt\lambda=2/3$, $\eta=1$, and $\lambda=4/9$; this corresponds to a bubble surface that falls from its 10\% to 90\% values over a distance of approximately $4.4$.
In this paper, the speed of the bubble walls is taken to 
be c, as might be expected for a very strong phase transition. It is known from previous studies of the 
electroweak phase transition that before the bubbles collide the walls reach a constant speed, where friction from the plasma and pressure inside the bubbles balance, and that the bubble wall speed 
is definitely less than c [14].   It was found in \cite{ahonen} that, for the case of Abelian bubble collisions, the resulting magnetic fields decrease in strength with decreasing wall 
speed, and we expect our fields to scale similarly. This issue was discussed further in \cite{mv}, and we plan to develop future calculations with a realistic wall 
speed.  We assume that the bubbles nucleate at $T=T_C=166$ GeV as in Ref.~\cite{mv}.

\begin{figure}[h!]
\includegraphics{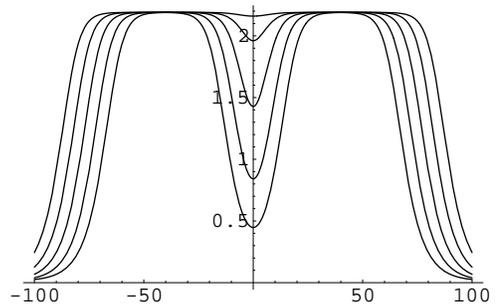}
\vspace{4.5cm}
\caption{Parameterized form for the scalar field of two colliding 
bubbles expending about their own centers and coalescing.  
The field is shown along the z-axis for the first few time steps for 
bubbles larger than the ones we have used to calculate 
the magnetic 
field.}
\label{wcp2}
\end{figure}

For two colliding bubbles, rather than solving (\ref{eomf}) directly,  we choose a simple parameterized form for the scalar fields
such that the bubble walls will expand smoothly, and that the scalar field will remain exactly
constant throughout the bubbles in the broken symmetry phase. Graphs of this
parameterization are shown in Fig.\ref{wcp2}. For calculating $m(x)$ from the scalar field we take into account that there is also an average scalar field from the other bubbles in the medium, which in this work is assumed to have a magnitude of 10 \% of the scalar field in the interior of a bubble.

\subsection{Current and Magnetic Field}

In cylindrical coordinates, using the form as before
\begin{equation}
j_{\nu} = \left\{ 
\begin{array}{ll}
         j_0, & \mbox{$\nu$=$0$};\\
         jx_{\nu} & \mbox{$\nu$=$1,2$};\\
        j_z & \mbox{$\nu$=$3$}.
\end{array} 
\right. 
\end{equation}
the current (\ref{current}) can  be written as $(\nu=0)$:

\begin{eqnarray}
4\pi j_0 &=& k\epsilon^{ab3}[
w_0^b(\partial\cdot w^a) +r^2 w^a \frac{\partial w^{b}}{\partial t} \\ \nonumber 
& &+w_z^a \frac{\partial w_z^{b}}{\partial t}+w_0^a \frac{\partial w_0^{b}}{\partial t} \\ \nonumber
&&+2w^a r \frac{\partial w_0^{b}}{\partial r}-2w_z^a \frac{\partial w_0^{b}}{\partial z}]
\end{eqnarray}
and for $(\nu=1,2)$:

\begin{eqnarray}
4\pi j &=& k\epsilon^{ab3}[
w^b(\partial\cdot w^a)+\frac{w_0^a}{r} \frac{\partial w^{0b}}{\partial r}\\ \nonumber
& &
-\frac{w_z^a}{r} \frac{\partial w_z^{b}}{\partial r} +2 w_0^a \frac{\partial w^{b}}{\partial t} \\ \nonumber 
&&+ w^a r \frac{\partial w^{b}}{\partial r}
-2w_z^a \frac{\partial w^{b}}{\partial z}]
\end{eqnarray}
and for $(\nu=3)$:

\begin{eqnarray}
4\pi j_z &=& k\epsilon^{ab3}[
w_z^b(\partial\cdot w^a)-w_0^a \frac{\partial w^{0b}}{\partial z}\\ \nonumber
& &+r^2 w^a \frac{\partial w^{b}}{\partial z} 
+2w_0^a \frac{\partial w_z^{b}}{\partial t}\\ \nonumber
&&+2w^a r \frac{\partial w_z^{b}}{\partial r}
-w_z^a \frac{\partial w_z^{b}}{\partial z}]
\end{eqnarray}

Having obtained the current, we can now calculate the magnetic fields directly from
the Maxwell equations
\begin{equation}
\partial^{\mu}F_{\mu\nu}=j_\nu~,
\end{equation}
which may be written in terms of the vector potential $A_{\mu}$ as
\begin{equation}
\partial^2 A_\nu-\partial_{\nu}(\partial^{\mu}A_{\mu})=j_{\nu}~.
\end{equation}
Working in the axial gauge
\begin{equation}
A_z=0
\end{equation}
the $(\nu=z)$ equation becomes
\begin{equation}
-\partial_z(\partial_{\mu}A^{\mu})=j_z~,
\end{equation}
which specifies the divergence of $A_{\mu}$ as
\begin{equation}
\partial_{\mu}A^{\mu}=-\int^z_{-\infty} j_z dz'~.
\end{equation}

Taking the form of the vector potential in cylindrical coordinates to be
\begin{equation}
A_{\nu} = \left\{ 
\begin{array}{ll}
         a_0, & \mbox{$\nu$=$0$};\\
         ax_{\nu} & \mbox{$\nu$=$1,2$};\\
        a_z & \mbox{$\nu$=$3$}~,
\end{array} 
\right. 
\end{equation}
the equation for the vector potential $a(x)$ becomes

\begin{multline}
\label{snooker}
{\frac{\partial^2 a}{\partial t^2}-\frac{\partial^2 a}{\partial r^2}
-\frac{3}{r}\frac{\partial a}{\partial r}-\frac{\partial^2 a}{\partial 
z^2}}\\
+\frac{1}{r}\frac{\partial}{\partial r}(-\int^z_{-\infty} j_z dz') =  j
\end{multline}

Due to the axial symmetry of the collision, it can be easily shown that the only nonvanishing component of the magnetic field is $B_\phi$,
given by
\begin{equation}
\label{bphi}
B_\phi = -r\frac{\partial a}{\partial z}~.
\end{equation}
The quantity $B^\phi$ may be found directly from Maxwell's equations by taking the derivative of (\ref{snooker}) with respect to $z$
and using equation (\ref{bphi}),
\beq
\label{maxazb}
&&(\frac{\partial^2}{\partial t^2} -r\frac{\partial^2}{\partial
r^2}\frac{1}{r} -3\frac{\partial}{\partial r} \frac{1}{r}
-\frac{\partial^2}{\partial z^2} ) B^\phi \nonumber \\
  &=& 
4\pi \frac{\partial j_z (x)}{\partial r} + 4\pi r\frac{ \partial j 
(x)}{\partial z} ~.
\eeq


As a preliminary result, we solve (\ref{maxazb}) on a 
coarse grid using an interpolating function for the current. Equations (\ref{eom1},\ref{eom2},\ref{eom3}) were solved for $t=0$ to $30$, $r=.01$ to $60$ and $z=-80$ to $80$, and values for the current were calculated on a grid with a step size of $2$ in the $t$, $r$ and $z$ directions. {\em Mathematica}~\cite{MATH} was then used to construct a polynomial interpolating function used as the current in Eq. (\ref{maxazb}).



\subsection{Numerical results and comparison to previous work}

Here we show the magnetic field with the boundary conditions in Fig.~\ref{subdpic} and compare it to analogous results calculated from the theory of Ref.~\cite{mv}, plotting the fields at comparable intervals $\delta t$ following the onset
of the collision at $t_0=0$.

\begin{figure}[tbh!]
\includegraphics{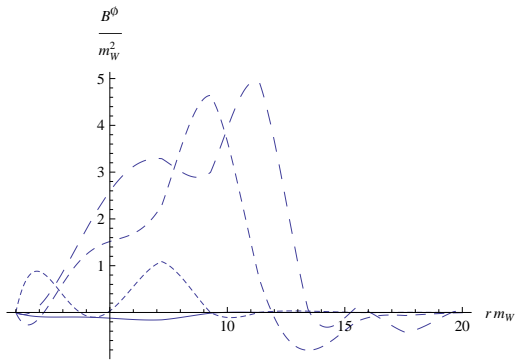}
\vspace{5cm}
\caption{ Magnetic field calculated in this work in 
the transverse direction 
at $z$=0 for times  $t=$5 (solid curve), 10 
(short dash curve), 15 (medium dash curve) and, 
20 (long dash curve) in units where the w-boson mass 
$m_W=1$. The magnetic field can be seen to be moving 
away from $r=0$, and increasing in magnitude, as $t$ increases.}
\label{magpic2}
 \end{figure}

\begin{figure}[tbh]
\includegraphics{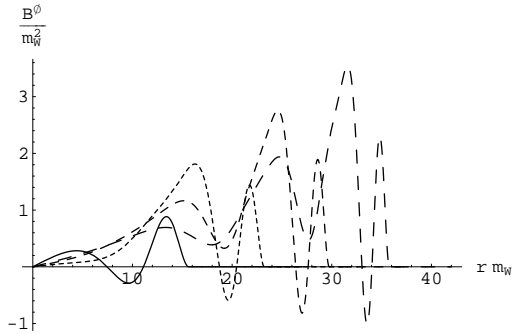}
\vspace{5cm}
\caption{Magnetic field calculated in \cite{mv} in the 
transverse direction at 
$z$=0 for times  $t=$5 (solid curve), 10 (short dash curve), 
15 (medium dash curve) and, 20 (long dash curve) 
in units where the w-boson mass $m_W=1$.}
\label{oldmagpic2}
 \end{figure}

The corresponding fields calculated from Ref.~\cite{mv} are shown in Fig.~\ref{oldmagpic2}. Bubble surface dynamics seems to produce fields somewhat larger in magnitude, and it is expected that bubble walls of even smaller surface thickness might grow even larger~\cite{stevens2}. The field in Fig.~\ref{magpic2} can be seen to be more concentrated near the center of the bubbles, and for this reason will have a smaller scale at the completion of the  phase transition.  
However, since the rate at which $w_z$ expands relative to the scalar field depends on the choice of $\partial w_z(t_0)/\partial t$, it would be interesting to quantify in future work the extent to which the scale and magnitude of the magnetic field might increase with a choice different from the one made on this paper, $\partial w_z(t_0)/\partial t =0$.

We have not attempted to determine the present day magnetic fields
that are seeded by our fields generated during the EWPT since this is a
complicated problem of plasma physics that has been studied
extensively elsewhere.  
The most recent of these~\cite{jed} show the importance of helicity and supports the possibility that
galactic cluster magnetic fields may be entirely primordial in origin.

\section{Summary and Conclusion}

We have shown how bubble surface dynamics affect magnetic seed field creation in collisions of bubbles in primordial first-order electroweak phase transition by extending the study in Ref.~\cite{mv} to treat, for the first time, the case of collisions of bubbles with walls of finite thickness. 
By working in the linear regime of gentle collisions, we are able to decouple the coupled and highly nonlinear partial differential equations that describe the evolution of the scalar and $w$ fields equations appearing in the Lagrangian of an appropriate extension of the Standard Model.  This simplifies the equations, making the influence of the surface dynamics relatively easy to study numerically. We find results in qualitative agreement with the previous studies, but allowing for the possibility that the magnetic seed fields could be even larger for sufficiently weak first-order phase transitions in which
bubbles occur with walls of even smaller thickness. Our work thus leaves open the possibility that the magnitude of the magnetic seed fields could be even larger than those that we find, making them an even more likely candidate for the origin of observed galactic and extra-galactic magnetic fields.  In this case the observation of these present day fields hold important clues to the form of the extension of the Standard Model, a subject of intense interest in physics because of its fundamental importance.

\section{Acknowledgements}

Dr. Stevens would like to acknowledge the NASA West Virginia Space Grant Consortium for partial support of this research through a Research Initiation Grant.  MBJ and TS thank Los Alamos National Laboratory for its support.


\end{document}